\documentclass[preprint,aps,12pt,preprintnumbers,eqsecnum,nofootinbib]{revtex4}
\usepackage{xcolor}
\usepackage{graphicx}
\usepackage{subcaption}
\usepackage{epstopdf} 
\usepackage{braket}
\usepackage{amssymb,amsmath,amsfonts,comment,latexsym,graphicx,epstopdf,float}
\usepackage{color}
\usepackage{fancyvrb}
\usepackage{chngcntr}
\usepackage{etoolbox}
\usepackage{array}
\usepackage{slashed}

\usepackage[colorlinks=true,citecolor=blue,linkcolor=red,filecolor=cyan,urlcolor=magenta]{hyperref}

\newcommand{\be}{\begin{equation}}
\newcommand{\ee}{\end{equation}}
\newcommand{\ba}{\begin{eqnarray}}
\newcommand{\ea}{\end{eqnarray}}

\newcommand{\grts}{\raise.3ex\hbox{$>$\kern-.75em\lower1ex\hbox{$\sim$}}}
\newcommand{\lets}{\raise.3ex\hbox{$<$\kern-.75em\lower1ex\hbox{$\sim$}}}

\newcommand{\dd}{\text{d}}

{\catcode`\|=\active\gdef\Braket#1{\left<\mathcode`\|"8000\let|\bravert 
{#1}\right>}}

\def\bravert{\egroup\,\vrule\,\bgroup}

\usepackage{color}
\usepackage{amssymb,amsmath,amsfonts}
\usepackage{epstopdf} 
\usepackage{braket}

\usepackage[export]{adjustbox} 
\newcommand\BubbleDiagram{ \makebox{\raisebox{-0.7cm}{\includegraphics{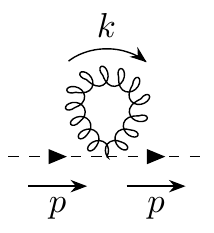}~}} }
\newcommand\RainbowDiagram{ \makebox{\raisebox{-0.8cm}{\includegraphics{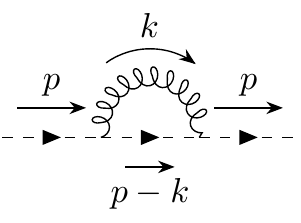}}} }

\newcommand\OneGluonVertex{ \makebox{\raisebox{-0.75cm}{\includegraphics{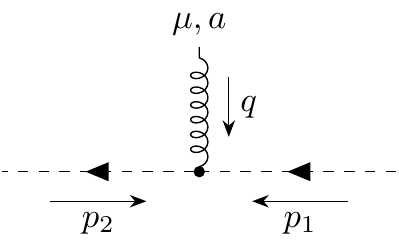}~}} }
\newcommand\TwoGluonVertex{ \makebox{\raisebox{-0.75cm}{\includegraphics{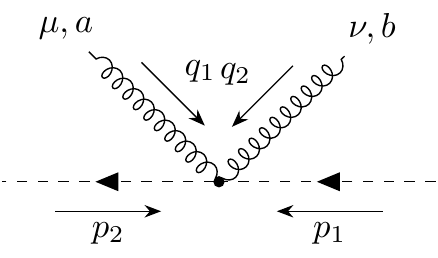}~}} }

\newcommand\ThreeGluonVertex{ \makebox{\raisebox{-0.75cm}{\includegraphics{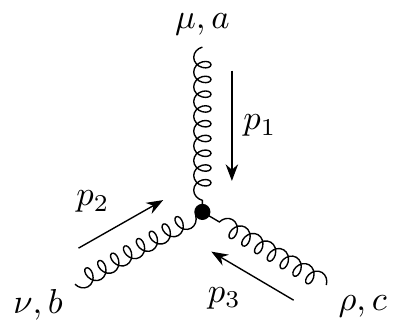}~}} }
\newcommand\FourGluonVertex{ \makebox{\raisebox{-1.75cm}{\includegraphics{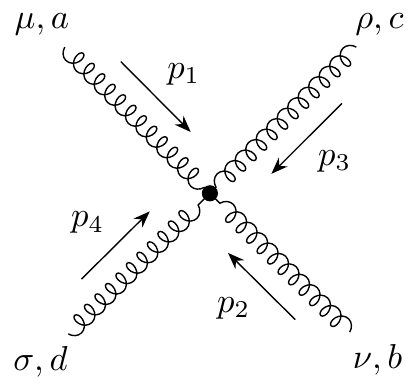}~}} }

\begin{document}
%
%
\title{\vspace*{0.5in} 
On asymptotic nonlocality in non-Abelian gauge theories
\vskip 0.1in}
\author{Jens Boos}\email[]{jboos@wm.edu}
\author{Christopher D. Carone}\email[]{cdcaro@wm.edu}
\affiliation{High Energy Theory Group, Department of Physics,
William \& Mary, Williamsburg, VA 23187-8795, USA}
%
%
\date{February 14, 2022}
\begin{abstract}
Asymptotically nonlocal field theories represent a sequence of higher-derivative theories whose limit point is a ghost-free, infinite-derivative theory.   Here, we extend previous work on pure scalar and Abelian gauge theories to asymptotically nonlocal non-Abelian theories.   In particular, we confirm that there is a limit in which the Lee-Wick spectrum can be decoupled, but where the hierarchy problem is resolved via an emergent nonlocal scale that regulates loop diagrams and that is hierarchically smaller than the lightest 
Lee-Wick resonance.
\end{abstract}

\maketitle

\section{Introduction}\label{sec:intro}

A substantial literature exists on higher-derivative theories, including those with quadratic term that are modified by an operator involving finite or infinite numbers of derivatives \cite{Boos:2021chb,Boos:2021jih,Efimov:1967,Krasnikov:1987,Kuzmin:1989,Tomboulis:1997gg,Modesto:2011kw,Biswas:2011ar}. Consideration of such theories are well motivated given their promise of offering better convergence properties of loop amplitudes. In Refs.~\cite{Boos:2021chb,Boos:2021jih}, we defined a novel class of higher-derivative theories that represent a sequence whose limit point is a ghost-free, infinite-derivative theory.  A specific theory in this sequence with $N$ propagator poles for a given field is suitable for eliminating a scalar mass hierarchy problem if the Lee-Wick partners are comparable to the scale that one wishes to keep hierarchically below the cutoff of the theory.  This is the way things work in the Lee-Wick Standard Model~\cite{Grinstein:2007mp}, where $N=2$, as well as generalizations to $N=3$~\cite{Carone:2008iw} that have been discussed in the literature.   What is interesting about asymptotically nonlocal theories is that there is also a large $N$ limit in which the Lee-Wick particles become 
heavy (and approach degeneracy) but where the hierarchy problem is still resolved: loop diagrams are regulated in this limit by an emergent nonlocal scale, $M_\text{nl}$, that is hierarchically smaller that the mass of the lightest Lee-Wick resonance, $m_1$:
\begin{align}
M_\text{nl}^2 \sim {\cal O}\left(\frac{m_1^2}{N}\right) \, .
\end{align}
The nonlocal scale does not appear as a fundamental parameter in the Lagrangian.  The number of propagator poles provides a 
parametric origin for the large separation between the regulator scale and the heavy particle masses.  This allows  for the stabilization of a 
hierarchy between light scalar masses and the heavier mass scales in the theory.

To understand why an emergent scale arises that regulates loop diagrams, it is useful to recall the toy model of real scalars discussed in
Ref.~\cite{Boos:2021chb}, which was written initially in the form
\begin{equation}
{\cal L}_N = -\frac{1}{2} \, \phi_1 \Box \phi_N - V(\phi_1) - \sum_{j=1}^{N-1} \chi_j \, \left[ \Box \phi_j - (\phi_{j+1}-\phi_j)/a_j^2\right] \,\,\, .
\label{eq:start}
\end{equation}
Here the constants $a_j$ have units of length, and a possible prefactor multiplying the terms in the sum has been set to one by
a rescaling of the $\chi_j$ fields.    As discussed in Refs.~\cite{Boos:2021chb,Boos:2021jih}, constraints are obtained when one integrates over 
the $\chi_j$ in the generating functional for the theory: 
\begin{equation}
\Box \phi_j -(\phi_{j+1}-\phi_j)/a_j^2=0 \,\,\, ,\,\,\,\,\,\, \mbox{ for }j=1 \dots N-1.
\label{eq:recursive}
\end{equation}
This allows one to eliminate the $\phi_j$, for $j=2 \ldots N$.  It follows that
\begin{equation}
\phi_N = \left[\prod_{j=1}^{N-1} \left(1+\frac{\ell_j^2 \Box}{N-1} \right) \right] \phi_1 \,\,\, ,
\end{equation}
where $\ell_j^2 \equiv (N-1) \, a_j^2$, so that Eq.~(\ref{eq:start}) may be re-expressed as
\begin{equation}
\label{eq:hd-lagrangian}
{\cal L}_N = - \frac{1}{2} \phi_1 \Box \left[\prod_{j=1}^{N-1} \left(1+\frac{\ell_j^2 \Box}{N-1} \right) \right] \phi_1 - V(\phi_1)  \,\,\, .
\end{equation}
Alternatively, one may think of Eq.~(\ref{eq:hd-lagrangian}) as the fundamental Lagrangian for the theory.
In the limit where $N$ approaches infinity and the $\ell_j$ simultaneously approach a common value $\ell$, 
Eq.~(\ref{eq:hd-lagrangian}) approaches the asymptotic form
\begin{equation}
{\cal L}_\infty = -\frac{1}{2} \, \phi_1 \, \Box \, e^{\ell^2 \Box} \, \phi_1 - V(\phi_1) \,\, .
\label{eq:Linf}
\end{equation}
This nonlocal Lagrangian has been studied extensively in the literature (see Ref.~\cite{Buoninfante:2018mre,Boos:2020qgg} for applications in quantum field theory, gravity, 
and additional historical references). What is significant here is that the scale $\ell$ serves as a regulator of loop diagrams. It was confirmed by explicit calculations in 
Ref.~\cite{Boos:2021chb} that the same is true when $N$ is large but finite; the same behavior was found in the Abelian gauge theories presented in Ref.~\cite{Boos:2021jih}. 
The finite-$N$ formulation is also useful in that it allows one to avoid some of the complications related to unitarity that are inherent to the limiting theory, where the propagator 
involves a non-polynomial entire function of the momentum~\cite{Carone:2016eyp,Pius:2016jsl,Briscese:2018oyx,Briscese:2021mob,Koshelev:2021orf}.

It is worth noting that each choice of $N$ in Eq.~(\ref{eq:hd-lagrangian}) corresponds to a distinct theory with different kinetic terms, each varying from 
the exponential form that they approximate.  Nonetheless, at large-but-finite $N$, the regulator scale set by $\ell$ emerges.  Moreover, 
it was shown in Ref.~\cite{Boos:2021chb} that the same result is obtained numerically when one varies the assumed form of the mass spectrum at fixed 
$N$.  These observations suggest that the emergence of the nonlocal scale does not depend sensitively on the exponential form of 
the differential operator that appears in the limiting theory, Eq.~(\ref{eq:Linf}), but rather on the requirement that some entire function emerges 
that accounts for the desired ultraviolet momentum suppression in the propagator.  This statement could be tested further by 
considering constructions that lead to an appropriate differential operator that is not exponential in form; however, one then loses the 
simple auxiliary field construction in  Eq.~(\ref{eq:start}), as well as the relatively tractable higher-derivative loop calculations (presented later) 
that are facilitated by the present assumptions.  We therefore defer consideration of other functional forms for this differential operator 
to future work.

The higher-derivative modifications of $\phi^4$ theory in Ref.~\cite{Boos:2021chb} and of the Abelian gauge theory in Ref.~\cite{Boos:2021jih} 
affected only the propagators of these theories, so that all amplitudes were finite quantities.   Since these finite theories are described in the
asymptotically nonlocal limit by a  Lagrangian involving only light particle masses and one dimensionful scale $\ell$, as in Eq.~(\ref{eq:Linf}),  one can make a 
convincing dimensional argument that the nonlocal scale must regulate amplitudes at any loop order in this limit~\cite{Boos:2021chb,Boos:2021jih}.  The 
situation in non-Abelian gauge theories is less clear, due to an important qualitative difference: gauge invariance implies that
higher-derivative quadratic terms are accompanied by derivative interaction terms as well.  These interaction terms grow with energy so that we can no longer
conclude immediately that we have a finite large-$N$ theory; a cutoff $\Lambda$ introduces an additional dimensionful scale, potentially spoiling the previous
argument.  Thus, it is well motivated to take a closer look at the loop corrections to scalar masses in non-Abelian gauge theories to determine whether
the asymptotically nonlocal solution to the hierarchy problem found in the theories of Refs.~\cite{Boos:2021chb,Boos:2021jih}  is still obtained.  
We do so in this paper and report positive results.

This paper is organized as follows.  In Sec.~\ref{sec:hdform}, we define an asymptotically nonlocal non-Abelian gauge theory of a complex scalar field, in
higher-derivative form; we determine the relevant Feynman rules and obtain an expression for the superficial degree of divergence of the theory.  In 
Sec.~\ref{sec:mrenorm}, we show by explicit computation that the corrections to the complex scalar mass remain finite in this theory, despite the presence of
derivative interaction terms, and that the asymptotically nonlocal behavior found in the scalar and Abelian gauge theories of Refs.~\cite{Boos:2021chb,Boos:2021jih} 
is also obtained here.  In Sec.~\ref{sec:discuss}, we summarize our results and discuss the relationship to the preliminary discussion on non-Abelian theories 
given in Ref.~\cite{Boos:2021jih}, where the scalar sector was written in Lee-Wick form ({\em i.e.}, the form in which higher-derivative terms are absent).  For completeness, we provide an appendix with the Feynman rules for
the pure gauge sector of the theory, which may be useful for future phenomenological studies.

\section{Higher-derivative Yang-Mills theory} \label{sec:hdform} 

In our previous considerations of asymptotically nonlocal $\phi^4$ theory~\cite{Boos:2021chb} and Abelian gauge theory~\cite{Boos:2021jih}, we provided a 
higher-derivative formulation of each theory, and also an alternative in which higher-derivative quadratic terms are eliminated in favor of auxiliary fields, for arbitrary values 
of $N$.   Both give equivalent physical results. As we noted in  Ref.~\cite{Boos:2021jih}, it is technically difficult to construct a simple, gauge-invariant auxiliary-field 
formulation for asymptotically nonlocal non-Abelian gauge theories with $N$ arbitrary.  Moreover, gauge-boson self-interactions are encoded simply in the 
higher-derivative description, avoiding potentially complicated interaction terms between towers of Lee-Wick resonances that would appear in the alternative 
approach (see, for example, the form of those interactions in an $N=3$ theory in Ref.~\cite{Carone:2008iw}). Fortunately, the higher-derivative formulation is 
sufficient to address the issues raised in Sec.~\ref{sec:intro}, and we choose to work in that framework henceforth.   

\subsection{Lagrangian}

We focus our attention on the following higher-derivative Lagrangian:\footnote{Alternatively, we could have chosen the operator $f(\underline{\Box} + m_\phi^2)$ in the scalar sector, so that the propagator has its canonical residue when $p^2=m_\phi^2$.  This has no effect on our conclusions.}
\begin{align} 
\label{eq:lagrangian}
\mathcal{L} &= -\frac12 \text{Tr} \, F{}_{\mu\nu} f(\underline{\Box}) F{}^{\mu\nu} - \phi^\ast ( \underline{\Box} + m_\phi^2 ) f(\underline{\Box}) \phi 
-V(\phi)\, ,
\end{align}  
where the field strength tensor is given by
\begin{align}
F{}_{\mu\nu} = \partial{}_\mu A{}_\nu - \partial{}_\nu A{}_\mu - ig \, [A{}_\mu, A{}_\nu] \, .
\end{align}
We employ matrix notation $A_\mu \equiv A{}_\mu^a T^a$ and $F{}_{\mu\nu} \equiv F{}^a_{\mu\nu} T^a$, where $T^a$ denotes the generators of the gauge group and summation over the repeated Lie algebra indices is assumed. The higher-derivative operator
\begin{align}
f(\underline{\Box}) \equiv \prod\limits_{j=1}^{N-1} ( 1 + a_j^2 \underline{\Box} ) \, , 
\end{align}
is constructed from gauge-covariant derivatives and we assume that the constants $a_j > 0$ are non-degenerate.   In our notation  $\Box \equiv \partial{}^\alpha \partial{}_\alpha$ and $\underline{\Box} \equiv D{}^\alpha D{}_\alpha$, where
\begin{equation}
D{}_\mu \phi \equiv \left( \partial_\mu - i g A{}_\mu \right) \phi \, ,
\end{equation}
for a field $\phi$ in the fundamental representation, and
\begin{equation}
D{}_\mu X \equiv \partial{}_\mu X - ig \, [A{}_\mu, X] \, ,
\end{equation}
for an adjoint field $X$.   While in principle there could be different higher-derivative operators in the gauge sector and the matter sector, we assume for simplicity
that they coincide.   We also define the quadratic Casimir operator
\begin{align}
C_2 \, \delta^i_j\equiv (T^a)^i_k (T^a)^k_j \, ,
\end{align}
with sums on repeated indices implied.  The numerical value of $C_2$ depends on the gauge group and field representation under consideration.

\subsection{Feynman rules} \label{subsec:feyn}
Let us now develop perturbation theory to study the physical content of the Lagrangian \eqref{eq:lagrangian}. The higher-derivative scalar propagator is given by
\begin{align}
D(p) = \frac{i}{p^2-m_\phi^2} \frac{1}{f(-p^2)} \, .
\label{eq:D}
\end{align}
In order to find the gauge propagator one may follow the usual local gauge-fixing procedure to arrive at  
\begin{align}
D_{\mu\nu}^{ab}(p) = -i\frac{\eta{}_{\mu\nu} - \frac{p_\mu p_\nu}{p^2} \left[1 - \xi f(-p^2) \right]}{p^2 f(-p^2)}  \delta^{ab} \, .
\label{eq:DD}
\end{align}
We will discuss pole prescriptions in Sec.~\ref{sec:mrenorm}.   The Lagrangian \eqref{eq:lagrangian} gives rise to vertices with two scalars 
and up to $2N$ gluons.  For the calculation presented in Sec.~\ref{sec:mrenorm}, we will need the scalar-gluon vertex for one and two gluons, respectively. For finite $N$, they are given by
\begin{align}
V_{1\text{g}} &\equiv \OneGluonVertex = i g T^a (p_1 - p_2)^\mu \left[ f_1^N(p_1) - (p_2^2-m_\phi^2)f^N_2(p_1, p_2) \right] \, , \label{eq:v1g} \\
V_{2\text{g}} &\equiv \TwoGluonVertex = ig^2 T^a T^b \bigg\{ \eta_{\mu\nu}\left[ f_1^N(p_1) - (p_2^2-m_\phi^2)f_2^N(p_1, p_2) \right] \label{eq:v2g}\\
&\hspace{20pt} + (2p_2 + q_1)_\mu (2p_1 + q_2)_\nu \left[ f_2^N(p_1,p_1+q_2) - (p_2^2-m_\phi^2) f_3^N(p_1, p_2, p_1+q_2) \right] \bigg \} \nonumber \\
&\hspace{20pt} + \big[ (q_1, \mu, a) \leftrightarrow (q_2, \nu, b) \big] \, , \nonumber
\end{align}
where we defined the abbreviations\footnote{We follow the convention that $\sum_{k=j}^n \equiv 0$ 
and $\prod_{k=j}^n \equiv 1$ if $j > n$. } 
\begin{align}
\begin{split}
f_1^N(p) &\equiv \prod\limits_{j=1}^{N-1} (1 - a_j^2 p^2) \, , \\
f_2^N(p_1, p_2) &\equiv \sum\limits_{k=1}^{N-1} a_k^2 \, \left[\prod\limits_{j=1}^{k-1} (1 - a_j^2 p_1^2)\right]\left[\prod\limits_{j=k+1}^{N-1} (1 - a_j^2 p_2^2)\right] \, , \\
f_3^N(p_1, p_2, p_3) &\equiv \sum\limits_{n=1}^{N-1}\sum\limits_{k=n+1}^{N-1} \! a_n^2 a_k^2 \, \left[ \prod\limits_{j=1}^{n-1} (1 - a_j^2 p_1^2) \right]\left[\prod\limits_{j=n+1}^{k-1} (1 - a_j^2 p_2^2) \right] \left[ \prod\limits_{j=k+1}^{N-1} (1 - a_j^2 p_3^2)\right] \, .
\end{split}
\end{align}
Note that these functions are completely symmetric under exchange of momentum arguments, and also only depend on the momenta's magnitudes. In the limiting case of $N\rightarrow\infty$ and $a_j^2\rightarrow \ell^2/(N-1)$ one can show
\begin{align}
\begin{split}
\lim\limits_{N\rightarrow\infty} f_1^N(p) &\equiv f_1(p) = e^{-\ell^2 p^2} \, , \\
\lim\limits_{N\rightarrow\infty} f_2^N(p) &\equiv f_2(p_1, p_2) = \frac{e^{-\ell^2 p_1^2}-e^{-\ell^2 p_2^2}}{p_2^2-p_1^2} \, , \\
\lim\limits_{N\rightarrow\infty} f_3^N(p) &\equiv f_3(p_1, p_2, p_3) = \frac{e^{-\ell^2p_1^2}}{(p_2^2-p_1^2)(p_3^2-p_1^2)} + \frac{e^{-\ell^2p_2^2}}{(p_1^2-p_2^2)(p_3^2-p_2^2)} + \frac{e^{-\ell^2p_3^2}}{(p_1^2-p_3^2)(p_2^2-p_3^2)} \, ,
\end{split}
\end{align}
where the right-hand side is generated by the following expression:
\begin{align}
f_n(p_1,\dots,p_n) \equiv \sum\limits_{j=1}^n e^{-\ell^2 p_j^2} \prod\limits_{\substack{k=1\\k\not=j}}^n \frac{1}{p_k^2-p_j^2} \, .
\end{align}
We note that the way in which the $a_j$ approach $\ell^2/(N-1)$ is not crucial; for example, the parametrization
\begin{equation}
a_j^2 = \left(1-\frac{j}{2 N{}^P}\right) \frac{\ell^2}{N} 
\label{eq:param}
\end{equation}
would achieve the desired limit with $P>1$.\footnote{In Refs.~\cite{Boos:2021chb,Boos:2021jih}, we used this parametrization with
$P=1$, which also achieves asymptotic nonlocality as the product in Eq.~(\ref{eq:hd-lagrangian}) still approaches an exponential up to $1/N$
corrections.}

\subsection{Superficial degree of divergence} \label{subsec:sdd}

In this subsection, we find an expression for the superficial degree of divergence of loop diagrams in the theory.  We are interested in 
diagrams that are potentially divergent, where  results may differ from the finite asymptotically nonlocal theories discussed in Refs.~\cite{Boos:2021chb,Boos:2021jih}.  
Our expression for the superficial degree of divergence will make clear why we focused on the Feynman rules given 
in Eqs.~(\ref{eq:v1g}) and (\ref{eq:v2g}).

The theory has three types of vertices which each have momentum dependence;  we let $p$ represent a generic momentum
and we work in the gauge where $\xi=0$. The $n$-gauge-boson self-interactions scale as $p^{2 N +2 -n}$ (to see this explicitly in the cases where $n=3$ and $4$, see the appendix); the $n'$-gauge boson-complex scalar vertices,  scale as $p^{2 N - n'}$; finally, the ghost vertices scale as $p^1$, as these arise exactly as in the local theory.   The gauge fields, complex scalar and the ghosts have propagators that scale as $p^{-2N}$, $p^{-2N}$ and $p^{-2}$, respectively.   Taking into account all these sources of momentum dependence, the superficial degree of divergence is given by
\begin{equation}
d=4\, L -2 \, N \, I -2\, I_{\rm gh} + \sum_n (2N+2-n) \, V_n + V_{\rm gh} -2 \, N \, I_{\rm s} + \sum_{n'} (2 N - n') \,V_{{\rm s} n'}  \,\,\, ,
\label{eq:sdd1}
\end{equation}
where $V_n$ is the number of pure-gauge vertices with $n$ gauge boson lines, $V_{{\rm s}n'}$ is the number of complex scalar vertices with $n'$ gauge boson lines, and $V_{\rm gh}$ are the number of ghost vertices; the number of loops is denoted by $L$, while the number of gauge, scalar and ghost internal lines are given by
$I$, $I_{\rm s}$ and $I_{\rm gh}$, respectively.    Four relations restrict the variables in Eq.~(\ref{eq:sdd1}):   The number of loop momenta is given by the number of 
internal line momenta that are not restricted by energy-momentum-conserving delta functions at the vertices, aside from overall energy-momentum conservation:
\begin{equation}
L = I + I_{\rm gh}+I_{\rm s} - \sum_n V_n -V_{\rm gh} - \sum_{n'} V_{{\rm s}n'} +1   \,\,\, .
\label{eq:con1}
\end{equation}
The number of ghost lines and complex scalar lines are separately conserved:
\begin{equation}
2 \, V_{\rm gh} = 2 \,I_{\rm gh} + E_{\rm gh}  \,\,\, ,
\label{eq:con2}
\end{equation}
\begin{equation}
2 \, \sum_{n'} V_{{\rm s}n'} = 2\, I_{\rm s}+E_{\rm s} \,\,\, .
\label{eq:con3}
\end{equation}
Here $E$, $E_{\rm gh}$, and $E_{\rm s}$  represent the number of external gauge-field, ghost and complex scalar lines, respectively.   Finally, all lines emanating from vertices must go somewhere:
\begin{equation}
\sum_n n \, V_n + 3 V_{\rm gh} + \sum_{n'} V_{{\rm s}n'} (n'+2) = 2\, I + 2\,I_{\rm gh} + 2\, I_{\rm s} +E +E_{\rm gh} + E_{\rm s} \,\,\, .
\label{eq:con4}
\end{equation}
It follows algebraically that one may use Eqs.~(\ref{eq:con1})--(\ref{eq:con4}) to rewrite Eq.~(\ref{eq:sdd1}) in the following useful form
\begin{equation}
d = 2N+2 - 2 (N-1) L - E - E_{\rm s} - N E_{\rm gh}  \,\,\, ,
\label{eq:sddf}
\end{equation}
which expresses the superficial degree of divergence in terms of the number of loops, external lines and N;  in the case were $N=2$, this expression agrees
with the one presented in Ref.~\cite{Grinstein:2007mp}.  Focusing on the scalar mass self-energy ($E=E_{\rm gh}=0$, $E_{\rm s}=2$), it is easy to confirm
that as $N$ becomes large (the limit of interest), all loop diagrams with $ L \geq 2$ become finite, while only the case where $L=1$ gives $d=2$ for any $N$.   
We therefore focus our attention on the one-loop scalar self-energy diagrams.   It was noted in the case of $N=2$ that this amplitude is more convergent than
would be indicated by Eq.~(\ref{eq:sddf})~\cite{Grinstein:2007mp}.  We will show in the next section that the amplitude is in fact finite in the asymptotically 
nonlocal limit, and is regulated by the emergent nonlocal energy scale.   

\section{Mass renormalization}\label{sec:mrenorm}  

The scalar self-energy evaluated at $p^2 = m_\phi^2$ determines the shift in the pole mass of the scalar that would have mass $m_\phi^2$ in the absence
of  radiative corrections.   Based on the conclusion of Sec.~\ref{subsec:sdd}, we 
consider the one-loop contributions to the self-energy in this section, with the goal of demonstrating two things:  (i) The sum of diagrams is finite, given our assumption 
that $N \geq 3$.  (ii) The amplitude in the large $N$ limit can be evaluated analytically, allowing one to confirm that the scale 
of the result is in fact set by the emergent nonlocal scale $1/\ell$.  Using the Feynman rules presented earlier, the two contributions to the self-energy,
$-i  \,M^2(p^2) = -i \,[M^2(p^2)_1+ M^2(p^2)_2]$, are given by    
\begin{align}
-i  \,M^2(p^2)_1 & \equiv \RainbowDiagram  \nonumber \\
&=  -4 \, g^2 C_2  \int \frac{d^4 k}{(2 \pi)^4}  \, \frac{p^2 - (p\cdot k)^2/k^2}{
k^2 f_1^N(k) [ (p-k)^2-m_\phi^2] f_1^N(p-k)} \label{eq:m1} \\
& \left[ f_1^N(p-k) - (p^2-m_\phi^2) f_2^N(p-k,p) \right]  \left[ f_1^N(p) - [(p-k)^2-m_\phi^2] f_2^N(p,p-k) \right] \,\,\,, \nonumber \\
-i  \,M^2(p^2)_2 & \equiv \BubbleDiagram  \nonumber \\
&=  g^2 C_2 \int \frac{d^4 k}{(2 \pi)^4}  \, \frac{1}{k^2 f_1^N(k)} \Big\{ 3 \left[f_1^N(p) - [p^2-m_\phi^2] f_2^N(p,p) \right] \label{eq:m2} \\
&\hspace{2em}-4 [p^2 - (p\cdot k)^2/k^2] \left[f_2^N(p,p-k) - [p^2-m_\phi^2] f_3^N(p,p-k,p)\right] \Big\} \,\,\, . \nonumber
\end{align}   
Notice that in our gauge choice, $\xi=0$, the gauge field propagators are proportional to $\eta_{\mu\nu} - k_\mu k_\nu /k^2$, which vanishes when contracted with $k^\mu$.  As 
a consequence, we can see by inspection that the most divergent terms in Eqs.~(\ref{eq:m1}) and (\ref{eq:m2}) taken separately are reduced from $d=2$ to $d=0$.\footnote{Physical quantities
like the shift in the pole mass are in fact gauge invariant, which was shown explicitly in the Abelian example of Ref.~\cite{Boos:2021jih} by keeping $\xi$ arbitrary and demonstrating the cancellation
of the $\xi$-dependent terms between the two diagrams that contribute to the amplitude.  In the interest of brevity, we do not repeat that exercise here.}  One may then 
show that the logarithmically divergent pieces cancel between $M^2(p^2)_1$ and $M^2(p^2)_2$.    For example, the vertices of the diagrams in Eqs.~(\ref{eq:m1}) and (\ref{eq:m2}) simplify in the limit
of large loop momenta $k$; with $p_1=p$ and $p_2 = -(p-k)$ in Eq.~(\ref{eq:v1g}) [or with $p_1=p-k$ and $p_2=-p$], and 
with $p_1=p$, $p_2=-p$ and $q_2=-k$ in Eq.~(\ref{eq:v2g}), the vertices become
\begin{align}
V_{1\text{g}} & \rightarrow i g \,T^a (2 p - k)^\mu (k^2)^{N-1}  \prod\limits_{j=1}^{N-1} (-a_j^2)   \,\,\, , \\
V_{2\text{g}} & \rightarrow i g^2 T^a T^b \, (2 p - k)^\mu (2 p - k)^\nu (k^2)^{N-2 } \prod\limits_{j=1}^{N-1} (-a_j^2) \,\,\, ,
\end{align}
in the limit that $k \rightarrow  \infty$.  One can then isolate the leading logarithmically divergent parts of Eqs.~(\ref{eq:m1}) and (\ref{eq:m2}) for $N \geq 3$; one finds
\begin{equation}
-i \,M^2(p^2)_1 =  +\, i \, M^2(p^2)_2 \rightarrow -\, 4 \, g^2 C_2 \int \frac{d^4 k}{(2 \pi)^4} \frac{p^2-(p\cdot k)^2/k^2}{k^4} \,\,\, ,
\end{equation}
revealing that the logarithmic divergences cancel.  Hence, the sum of Eqs.~(\ref{eq:m1}) and (\ref{eq:m2}) is finite;\footnote{Note that
there are terms in $M^2(p^2)_2$ that are subleading at large $N$, proportional to $\int \! d^4 k\,  (p^2)^{N-2}/(k^2)^N$; these become log divergent when $N=2$, consistent with the one-loop results in the 
Lee-Wick Standard Model~\cite{Grinstein:2007mp}.} we again have the situation obtained in the scalar and Abelian gauge theories
of Refs.~\cite{Boos:2021chb,Boos:2021jih}, where a dimensional argument is available suggesting that the result is set by the emergent nonlocal scale as the limiting theory is approached.  

We now verify this explicitly.   Since the limit of interest is one in which $N$ is large [and the Lee-Wick spectrum is becoming increasingly 
degenerate, as in Eq.~(\ref{eq:param})], we evaluate the self-energy at leading order in a $1/N$ expansion.  The 
$f_j^N$ will approach functions of exponentials in this limit; Wick rotation is done in the finite-$N$ theory, so there are no problems associated 
with the directions in the complex energy plane where the exponentials blow up; when exponentials are displayed in Minkowski-space expressions, they are a mnemonic for the finite-$N$ expressions that approach them, and serve to accurately approximate the result. In the 
finite-$N$ theory, we use the Lee-Wick pole prescription, which is identical to the Feynman prescription when the decay width of the
Lee-Wick resonances is neglected, as in our lowest-order calculation.   If we were to work at higher order,  the Lee-Wick poles become complex 
conjugate pairs as their widths are turned on\footnote{Heavy Lee-Wick gauge particles can decay, for example, to two light scalars.  We assume the potential in Eq.~(\ref{eq:lagrangian}) provides scalar self interactions that allow the heavy Lee-Wick scalars to decay as well.}  and shift away from the real $k^0$ axis; the Lee-Wick prescription requires deforming the contours around the poles so that they remain in the same relative position as in the 
Feynman prescription~\cite{LeeWick:1969}; one may then Wick rotate.  There is only an ambiguity when poles overlap and pinch a contour, a 
situation which requires an additional prescription to define the loop integral~\cite{Cutkosky:1969fq}.\footnote{It is worth noting that there are alternatives to the 
approach of Refs.~\cite{LeeWick:1969,Cutkosky:1969fq} that aim to address the classical instabilities of such 
higher-derivative theories.  See Refs.~\cite{Kaparulin:2014vpa,Kaparulin:2021kle} for discussion and an application to a non-Abelian model.}   However, for the situation we 
consider, where $p^2=m_\phi^2$ and all Lee-Wick poles much heavier, such pinching of contours does not occur.   (See, for example, the discussion 
in Sec. IV.B of Ref.~\cite{Grinstein:2008bg}.)

\begin{figure}[!htb]
\centering
\includegraphics[width=0.7\textwidth]{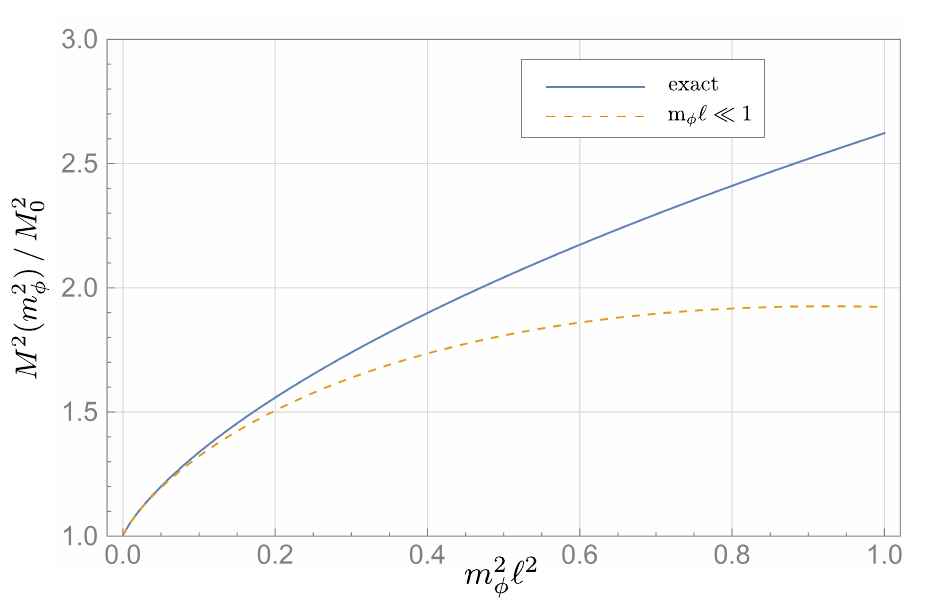}
\caption{Scalar self-energy at one loop, normalized to $M_0^2 \equiv 3 \, g^2 \, C_2 \, e^{-\ell^2 m_\phi^2}/[16\pi^2\ell^2]$, plotted as a function of $m_\phi^2\ell^2$ (solid line), and the approximation $m_\phi \ell \ll 1$ (dashed line).}
\label{fig:self-energy}
\end{figure}

The choice $p^2=m_\phi^2$ leads to a significant simplification in the form of the self-energy.  We find 
\begin{equation}
-i M^2(m_\phi^2) = g^2 C_2 \, e^{-\ell^2\, m_\phi^2} \int \frac{d^4 k}{(2 \pi)^4} \,e^{\ell^2 k^2}  \left[
\frac{3}{k^2} - \frac{4 \, m_\phi^2}{k^2 (k^2-2\, p \cdot k)}
+ \frac{4 \,(p \cdot k)^2}{k^4(k^2-2\, p \cdot k)} 
\right]  \,.
\label{eq:onshell}
\end{equation}
The momentum integration can be evaluated by first combining denominators, Wick rotating and then writing the result as a Euclidean Gaussian integral by means of a Schwinger parameter.  We find
\begin{equation}
-i M^2(m_\phi^2) = - i \, \frac{3 g^2}{16 \pi^2} \, \frac{1}{\ell^2} \, C_2\,  e^{-\ell^2 \, m_\phi^2} [1 + I_1(m_\phi^2 \ell^2)- I_2(m_\phi^2 \ell^2)] \,\,\, ,
\label{eq:fpform}
\end{equation}
where
\begin{equation}
I_1(z)  = \frac{4}{3} \, z \int_0^1 dx \int_0^\infty dy \, \frac{y}{(1+y)^2} \, e^{- x^2 y \, z}  \,\,\, ,
\end{equation}
and
\begin{equation}
I_2(z)  = \frac{2}{3}\, z \int_0^1 dx  \, (1-x)  \int_0^\infty dy \, \frac{y^2[1-2 x^2 (1+y) z]}{(1+y)^3} \, e^{- x^2 y\,  z} \,\,\, .
\end{equation}
Note that the three terms in Eq.~(\ref{eq:fpform}) correspond to the three terms in square brackets in Eq.~(\ref{eq:onshell}), which involve one, two and three denominator factors, 
respectively.  The functions $I_1$ and $I_2$ are positive and never exceed $\sim 1.7$ for $m_\phi^2 \ell^2$ between $0$ and $1$, consistent with the assumption that the scalar mass is below
the emergent nonlocal scale. The integrals can be performed analytically and we find
\begin{equation}
\label{eq:mphi-exact}
-i M^2(m_\phi^2) = -i \frac{3g^2}{16\pi^2} \frac{1}{\ell^2} C_2 e^{-\ell^2 m_\phi^2} \left[ 1 + F(m_\phi^2\ell^2) \right] \, ,
\end{equation}
where
\begin{equation}
F(z) = \frac16 \left[ (z^2-2z) e^z \text{Ei}_1(z) - z\right] + \frac{\sqrt{z}}{6} \left[ G{}^{22}_{23}\left( z \left| \begin{matrix} -\tfrac12, 1; \\ \tfrac32, \tfrac12; 0 \end{matrix} \right . \right) + 4 \, G{}^{22}_{23}\left( z \left| \begin{matrix} -\tfrac12, 1; \\ \tfrac12, \tfrac12; 0 \end{matrix} \right. \right) \right] \, .
\end{equation}
Here $G^{mn}_{pq}$ denotes the Meijer G-function \cite{Olver:2010} and $\text{Ei}_1(z) \equiv \int_1^\infty \dd t \, e^{-tz}/t$ is an exponential integral. For $m_\phi^2 \ell^2 \ll 1$ the expressions simplify, and we find
\begin{align} \label{eq:mphi-approx}
-i M^2(m_\phi^2) \approx - i \, \frac{3 g^2}{16 \pi^2} \, \frac{1}{\ell^2} \, C_2\,  e^{-\ell^2 \, m_\phi^2}\left[ 1 + \left( \frac32 - \gamma - \log m_\phi^2\ell^2 \right) m_\phi^2\ell^2 \right] + \mathcal{O}(m_\phi^4\ell^4) \, .
\end{align}
We note that this leading-order expression in $m_\phi^2\ell^2 \ll 1$ closely resembles the expression found in the Abelian case in Ref.~\cite{Boos:2021jih}; for a plot of the exact result, Eq.~\eqref{eq:mphi-exact}, and the small-$m_\phi$ result, Eq.~\eqref{eq:mphi-approx}, see Fig.~\ref{fig:self-energy}. Hence, the scalar pole mass does not receive a loop correction that exceeds the nonlocal scale $1/\ell$ in the limit that the heavy Lee-Wick particles are decoupled via a parameterization like the one in Eq.~(\ref{eq:param}), with $m_j = 1/a_j$.  This is the same behavior found the scalar and Abelian gauge theories considered in our previous work~\cite{Boos:2021chb,Boos:2021jih}.

Finally, we note that we could have defined the theory such that in the $N \rightarrow \infty$ limit, the exponential in the Lagrangian is $e^{-\ell^2 (p^2 - m_\phi^2)}$, rather than $e^{-\ell^2 \, p^2}$; this 
assures a canonically normalized tree-level propagator when $p^2=m_\phi^2$.  Proceeding in this way, the only change in Eq.~(\ref{eq:fpform}) is that the factor of $e^{-\ell^2 \, m_\phi^2}$ would 
not appear, and our qualitative conclusions would remain unchanged.

\section{Discussion}\label{sec:discuss} 

In this paper, we have tied up a loose end from Ref.~\cite{Boos:2021jih}, where asymptotically nonlocal non-Abelian gauge theories were defined in 
higher-derivative form, but a complete argument was not presented showing that corrections to the mass of a light scalar particle are set by the emergent 
nonlocal scale -- a scale that is hierarchically lower than the mass of the lightest Lee-Wick resonance.   Since non-Abelian theories in their higher-derivative 
form have derivative interaction terms yielding vertices that grow with momentum, these theories are qualitatively different from the $\phi^4$ and Abelian 
gauge theories considered in detail in Refs.~\cite{Boos:2021chb} and \cite{Boos:2021jih}, respectively.  In the present work, we first determined
the relevant Feynman rules for an asymptotically nonlocal non-Abelian theory in higher-derivative form, and computed the superficial degree of divergence for
an arbitrary loop diagram.  We showed that the only potentially divergent diagrams occur at one-loop as one approaches the nonlocal limiting theory.  Then, 
we evaluated these diagrams and found a nontrivial cancellation of divergent parts, so that the resulting self-energy is a finite 
quantity.  As a consequence, the dimensional arguments given in our earlier work apply and suggest that the scale of the radiative corrections at any loop 
order must be set by the emergent nonlocal scale $1/\ell^2$ in the desired limit, for lack of any alternative scale, aside from light particle squared masses.  We 
supported this conclusion by explicitly evaluating the scalar self-energy at lowest nontrivial order in perturbation theory in the asymptotically nonlocal limit; the result was found to be proportional to $1/\ell^2$, up to gauge couplings and expected numerical loop factors. 

It is worth noting that in Ref.~\cite{Boos:2021jih} it was argued that the complex scalar sector of the theory studied here could be written without higher 
derivatives via the use of auxiliary fields and appropriate field redefinitions, while no such trick was available for the non-Abelian gauge sector for a general 
asymptotically nonlocal theory, one with an arbitrary number of propagator poles.  This led to the observation that each scalar mass eigenstate in 
the Lee-Wick basis appeared to couple to the higher-derivative gauge sector like a local scalar field, and hence would have finite self-energies at one-loop.   
This statement, as applied to the lightest mass eigenstate,\footnote{It also implies that analogous statements hold for the heavier scalar mass 
eigenstates, if one were interested in the decoupled sector.} is consistent with results found in the present work, and serves as a nontrivial check of the calculation presented in Sec.~\ref{sec:mrenorm}.

The present work adds to the evidence that the asymptotically nonlocal standard model Lagrangian presented in Ref.~\cite{Boos:2021jih} will be regulated
by the emergent nonlocal scale, when Lee-Wick resonances remain far outside the reach of collider experiments.  Nevertheless, scattering amplitudes will be 
affected as energies approach the emergent nonlocal scale, which must not be far above the electroweak scale if the hierarchy problem is to be addressed. 
This implies phenomenological consequences at colliders, a topic we plan to address in future work.

\begin{acknowledgments}  
We thank the NSF for support under Grants PHY-1819575 and PHY-2112460.
\end{acknowledgments}

\appendix

\section{Gluon self-interactions}  \label{app:2-loop}
The pure gauge part of the Lagrangian is given by
\begin{align}
\mathcal{L}_\text{gauge} &= -\frac12 \text{Tr} F{}_{\mu\nu} f(\underline{\Box}) F{}^{\mu\nu} \, , \quad f(\underline{\Box}) = \prod\limits_{j=1}^{N-1}(1 + a_j^2 \underline{\Box}) \, .
\end{align}
Recall that $\text{Tr}(T^a T^b) = \tfrac12 \delta{}^{ab}$ and the expansion of the field strength tensor
\begin{align}
F{}^a_{\mu\nu} &= \partial{}_\mu A{}_\nu^a - \partial{}_\nu A{}_\mu^a + g f{}^{abc} A{}_\mu^b A{}_\nu^b \, .
\end{align}
The covariant $\Box$-operator acting on an adjoint field $X^a$ is given by
\begin{align}
\begin{split}
\underline{\Box} X^a &= \eta{}^{\rho\sigma} \left(\delta^{ac} \partial_\rho + g f^{abc} A{}^b_\rho \right)\left(\delta^{ce} \partial_\sigma + g f^{cde} A{}^d_\sigma \right) X^e \\
&= \left\{ \Box \delta{}^{ae} + g f^{abe} [(\partial^\rho  A{}^b_\rho) + 2 A{}^b_\rho \partial^\rho] + g^2 \eta{}^{\rho\sigma} f^{abc}f^{cde} A{}^b_\rho A{}^d_\sigma \right\} X^e \, .
\end{split}
\end{align}
Gluon self-interactions can come from within the field strength tensors as well as the $\underline{\Box}$-operators, and hence theories of the above form allow for gluon self-interactions of up to $2(N+1)$ gluons. In what follows, we limit our considerations to the three- and four-gluon vertices. To simplify our discussion, we shall refer to a vertex contribution as a $(k,l,m,n)$-term when it has $k$ gluons from the leftmost field strength tensor, $l$ and $m$ gluons from two separate $\underline{\Box}$ operators, and $n$ gluons from the rightmost field strength tensor. Here, $k,l,m,$ and $n$ can be either 0, 1, or 2.

Following these considerations, the three-gluon vertex takes the form
\begin{align} \label{eq:ap3g}
V_{3g} &\equiv \ThreeGluonVertex = Y{}_{abc}^{\mu\nu\rho}(p_1,p_2,p_3) + \text{all permutations}  \, , \\
Y{}_{abc}^{\mu\nu\rho}(p_1, p_2, p_3) &= -g f^{abc}\left[ f_1^N(p_1) p_1^\nu \eta{}^{\mu\rho} + \frac12 f_2^N(p_1,p_3)(p_1-p_3)^\nu (p_1 \!\cdot p_3 \eta{}^{\mu\rho} - p_1^\rho p_3^\mu) \right] \, .
\end{align}
The first term is a sum of $(2,0,0,1)$ and $(1,0,0,2)$, that is, it is generated purely by gluons from within field strength tensors, and the second term is a $(1,1,0,1)$-type where one gluon is taken from the sandwiched $\underline{\Box}$-operator. In the limiting case of ordinary Yang-Mills theory one has $f_1^N \equiv 1$ and $f_2^N \equiv 0$ and one recovers the usual three-gluon vertex.  Taking into account permutations, Eq.~(\ref{eq:ap3g}) has $3!$ terms.

The four-gluon vertex takes the form
\begin{align}\label{eq:ap4g}
V_{4g} &\equiv\FourGluonVertex = X{}_{abcd}^{\mu\nu\rho\sigma}(p_1,p_2,p_3,p_4) + \text{all permutations}  \, , \\
X{}_{abcd}^{\mu\nu\rho\sigma}(p_1, p_2, p_3,p_4) &= -ig^2 f^{abe}f^{cde} \bigg[ \frac14 f^N_1(p_3+p_4)\eta{}^{\mu\rho}\eta^{\nu\sigma} \nonumber \\
&\hspace{12pt}- f_2^N(p_1+p_2,p_4)(p_3+2p_4)^\rho p_4^\mu\eta{}^{\nu\sigma} \nonumber \\
&\hspace{12pt}-\frac12 f_2^N(p_1,p_4)\eta{}^{\nu\rho}(p_1\!\cdot p_4 \, \eta{}^{\mu\sigma} - p_1^\sigma p_4^\mu ) \nonumber \\
&\hspace{12pt}-\frac12 f_3^N(p_1, p_1+p_2, p_4) (2p_1+p_2)^\nu(p_3+2p_4)^\rho ( p_1\!\cdot p_4\eta{}^{\mu\sigma}-p_1^\sigma p_4^\mu ) \bigg] \, .
\end{align}
Here, the first term is of type $(2,0,0,2)$, the second term is a sum of $(2,1,0,1)$ and $(1,1,0,2)$, the third term is of type $(1,2,0,1)$, and the last term is generated by $(1,1,1,1)$. Again, in the limiting case of ordinary Yang-Mills theory one has $f_1^N \equiv 1$ and $f_2^N \equiv f^3_N \equiv 0$ and one recovers the usual four-gluon vertex. Taking into account permutations, Eq.~(\ref{eq:ap4g}) has $4!$ terms.

As our gauge-fixing procedure is identical to that of a local Yang-Mills theory, the ghost Feynman rules are unaffected, so we do not display them here; they can be found in standard references, for example Ref.~\cite{Peskin:1995ev}.

\end{document}